# Isotopic evolution of the inner Solar System inferred from molybdenum isotopes in meteorites

Fridolin Spitzer[1], Christoph Burkhardt[1], Gerrit Budde[1,2], Thomas S. Kruijer[3], Alessandro Morbidelli[4], and Thorsten Kleine[1]

[1]Institut für Planetologie, University of Münster, Wilhelm-Klemm-Str. 10, 48149 Münster, Germany; fridolin.spitzer@uni-muenster.de

[2]The Isotoparium, Division of Geological and Planetary Sciences, California Institute of Technology, 1200 E California Blvd, Pasadena, CA 91125, USA.

[3]Nuclear & Chemical Sciences Division, Lawrence Livermore National Laboratory, 7000 East Avenue (L-231), Livermore, CA 94550, USA.

[4]Laboratoire Lagrange, UMR7293, Université de Nice Sophia-Antipolis, CNRS, Observatoire de la Côte d'Azur, Boulevard de l'Observatoire, 06304 Nice Cedex 4, France

**ABSTRACT**

The fundamentally different isotopic compositions of non-carbonaceous (NC) and carbonaceous (CC) meteorites reveal the presence of two distinct reservoirs in the solar protoplanetary disk that were likely separated by Jupiter. However, the extent of material exchange between these reservoirs, and how this affected the composition of the inner disk are not known. Here we show that NC meteorites display broadly correlated isotopic variations for Mo, Ti, Cr, and Ni, indicating the addition of isotopically distinct material to the inner disk. The added material resembles bulk CC meteorites and Ca-Al-rich inclusions in terms of its enrichment in neutron-rich isotopes, but unlike the latter materials is also enriched in $s$-process nuclides. The comparison of the isotopic composition of NC meteorites with the accretion ages of their parent bodies reveals that the isotopic variations within the inner disk do not reflect a continuous compositional change through the addition of CC dust, indicating an efficient separation of the NC and CC reservoirs and limited exchange of material between the inner and outer disk. Instead, the isotopic variations among NC meteorites more likely record a rapidly changing composition of the disk during infall from the Sun's parental molecular cloud, where each planetesimal locks the instant composition of the disk when it forms. A corollary of this model is that late-formed planetesimals in the inner disk predominantly accreted from secondary dust that was produced by collisions among pre-existing NC planetesimals.

Isotopic evolution of the inner Solar System

1. INTRODUCTION

Nucleosynthetic isotope anomalies reveal a fundamental dichotomy between non-carbonaceous (NC) and carbonaceous (CC) meteorites (Budde et al. 2016; Warren 2011), which sample two spatially distinct reservoirs that coexisted in the early Solar System for several million years (Ma) (Kruijer et al. 2017). The prolonged spatial separation of the NC and CC reservoirs most likely reflects the formation of Jupiter, which acted as an efficient barrier against material exchange either by its growth itself (Kruijer et al. 2017; Morbidelli et al. 2016) or through a pressure maximum in the disk near the location where Jupiter later formed (Brasser & Mojzsis 2020). Although there is little doubt that the NC and CC reservoirs were spatially separated, the extent of material exchange between them remains poorly constrained. For example, the Jupiter barrier may have resulted in a filtering effect by which the inward drift of large grains was efficiently blocked, while smaller dust grains may have passed the barrier as part of the gas flow (Haugbølle et al. 2019; Weber et al. 2018). On this basis, it has been argued that the inner disk's isotopic composition was modified through the addition of inward drifting CC dust (Schiller et al. 2018, 2020). This interpretation, however, depends on the assumed starting composition of the inner disk, and on the unknown efficiency of the Jupiter barrier over time. Thus, understanding and quantifying any compositional evolution of the NC reservoir is of considerable interest, as it would allow reconstructing the structure and temporal evolution of the solar accretion disk, and the importance of Jupiter for separating the NC and CC reservoirs.

The NC-CC dichotomy has been identified for several elements and so far holds for all analyzed meteorites (Kleine et al. 2020; Kruijer et al. 2020). The dichotomy is particularly exploitable for Mo, which can distinguish between isotope variations arising from the heterogeneous distribution of matter produced by the *p*-, *s*-, and *r*-processes of stellar nucleosynthesis (Burkhardt et al. 2011). While there are large *s*-process Mo isotope variations among meteorites within both the NC and CC groups, all CC meteorites are characterized by an approximately constant *r*-process excess over NC meteorites (Budde et al. 2016; Kruijer et al. 2017; Poole et al. 2017; Worsham et al. 2017). This difference makes Mo isotopes ideally suited to identify any compositional change of the NC reservoir, because the continuous addition of CC dust to the NC reservoir would result in a characteristic isotopic shift of the NC composition towards an enrichment in *r*-process Mo isotopes over time. For identifying such a potential isotopic shift in the NC reservoir, iron meteorites are particularly important, because they derive from some of the earliest planetesimals formed within the NC reservoir (Kruijer et al. 2014) and, therefore, may have a distinctly different Mo isotopic composition compared to later-formed NC planetesimals.

Until now, no systematic Mo isotopic difference between early- and late-formed NC bodies has been identified (Budde et al. 2019). This might in part be due to the overall small Mo isotopic offset between the NC and CC reservoirs, but for iron meteorites may also reflect the modification of their Mo isotopic compositions by neutron capture reactions induced during cosmic ray exposure (CRE) (e.g., Worsham et al. 2017). Here, we employ Pt isotopes to quantify CRE-effects (Kruijer et al. 2013; Wittig et al. 2013) on Mo isotopes with unprecedented precision and use these data, combined with published data for other meteorite groups, to assess any compositional heterogeneity within the inner disk that may have arisen through material exchange between the NC and CC reservoirs.

## 2. MOLYBDENUM ISOTOPIC HETEROGENEITY OF THE INNER DISK

### 2.1. Correction of Cosmic Ray Effects

Several group IC, IIAB, IID, and IIIAB irons with variable CRE-effects on Pt isotopes were selected for this study. Except for the IID irons, only NC iron meteorites were selected, because this study aims to assess potential isotopic changes in inner disk composition. The IID irons were incorporated because one of them (Carbo) is among the most strongly irradiated irons known (Kruijer et al. 2013; Qin et al. 2015). Combined, the investigated samples include strongly and weakly irradiated irons, which makes it possible to precisely quantify CRE-effects on Mo isotopes.

Sample preparation and Mo and Pt isotope measurements followed previously established methods (Budde et al. 2019; Kruijer et al. 2013). Isotopic compositions were determined using a Thermo-Fisher Neptune *Plus* MC-ICP-MS at Münster and are reported in the $\varepsilon$-notation (parts-per-10,000 deviations from terrestrial standard values) after mass bias correction to the terrestrial $^{98}$Mo/$^{96}$Mo and $^{198}$Pt/$^{195}$Pt, respectively.

Samples of a given iron group show variable $\varepsilon^i$Mo values that correlate with $\varepsilon^{196}$Pt, indicating the presence of CRE-effects (Fig. 1). The $\varepsilon^i$Mo–$\varepsilon^{196}$Pt correlations are best defined for the IIAB and IID irons, both of which include samples with large CRE-effects. Nevertheless, the IC and IIIAB irons also display correlated $\varepsilon^i$Mo–$\varepsilon^{196}$Pt variations, and the $\varepsilon^i$Mo–$\varepsilon^{196}$Pt slopes are consistent for all groups. The pre-exposure $\varepsilon^i$Mo (i.e., unaffected by CRE) for each iron group can either be obtained from the intercept value at $\varepsilon^{196}$Pt = 0, or by individually correcting each sample to $\varepsilon^{196}$Pt = 0 and using the mean $\varepsilon^i$Mo–$\varepsilon^{196}$Pt slopes determined for the different iron groups. Both approaches yield identical results (Table 1) and provide precise pre-exposure $\varepsilon^i$Mo values for the IC, IIAB, IID, and IIIAB irons. Pre-exposure $\varepsilon^i$Mo values for the IIIE irons were calculated using previously published Mo and Pt isotopic data (Kruijer et al. 2017; Worsham et al. 2019) (Table 1). The pre-exposure $\varepsilon^i$Mo values of this study are a factor of ~5 more precise than previous results (Bermingham et al. 2018), and only for IIAB irons have values with comparable precision been previously reported (Yokoyama et al. 2019) (Table 1). Finally, pre-exposure $\varepsilon^i$Mo values for IVA irons were calculated by averaging data for samples having no CRE effects (Poole et al. 2017) and CRE-corrected data (Bermingham et al. 2018).

## 2.2. *Mo Isotope Variations among NC Meteorites*

In a diagram of $\varepsilon^{95}$Mo versus $\varepsilon^{94}$Mo, bulk meteorites plot along two distinct and approximately parallel lines, which were termed the NC (Non-Carbonaceous) and CC (Carbonaceous Chondrite) lines (Budde et al. 2016). The Mo isotopic variations along the NC- and CC-lines are predominantly governed by *s*-process variations, whereas the offset between the two lines reflects the characteristic *r*-process excess of the CC over the NC reservoir. For distinguishing between these different Mo isotope variations, it is useful to define $\Delta^{95}$Mo as the vertical deviation (in ppm) of a sample from an *s*-process mixing line passing through the origin (Budde et al. 2019):

$$\Delta^{95}\text{Mo} = (\varepsilon^{95}\text{Mo} - 0.596 \times \varepsilon^{94}\text{Mo}) \times 100. \tag{1}$$

The quantity 0.596 is the slope of *s*-process mixing lines defined by bulk samples and acid leachates from both NC and CC meteorites (Budde et al. 2019), which is indistinguishable from the slope obtained from mainstream presolar SiC grains (Stephan et al. 2019). Distinct $\Delta^{95}$Mo values, therefore, indicate Mo isotope heterogeneities unrelated to pure *s*-process variations.

The precise pre-exposure $\varepsilon^{95}$Mo and $\varepsilon^{94}$Mo values from this study reveal that some NC irons plot below the NC-line (Fig. 2), and have slightly lower $\Delta^{95}$Mo than the characteristic NC value ($\Delta^{95}$Mo = –9±2; Budde et al. 2019) (Table 1). Linear regression of available $\varepsilon^{95}$Mo and $\varepsilon^{94}$Mo data for NC meteorites (Table S1), including the precise data for NC irons from this study, yields a slope of 0.528±0.045 (MSWD = 0.85), which is shallower than the slope of the CC-line and the characteristic slope of a pure *s*-process mixing line (Fig. 2). Including leachate data for NC chondrites (Budde et al. 2019) results in a steeper slope (m = 0.595±0.011), which is consistent with that of the CC-line and pure *s*-process variations. However, the higher MSWD of 1.6 for this regression is above the upper acceptable limit of 1.45 for *N* = 41 (Wendt & Carl 1991), indicating additional scatter outside the analytical uncertainties. The $\varepsilon^{95}$Mo–$\varepsilon^{94}$Mo slope of bulk NC meteorites, therefore, is shallower than the predicted slope of a pure *s*-process mixing line. This results in a weak inverse correlation of $\Delta^{95}$Mo with $\varepsilon^{94}$Mo (Fig. 2) and indicates that the Mo isotope variations among NC meteorites do not solely reflect *s*-process but also additional *r*-process variations.

## 3. ISOTOPIC EVOLUTION OF THE INNER SOLAR SYSTEM

### 3.1. Mixing Trends in the NC Reservoir

The $\Delta^{95}$Mo and $\varepsilon^{94}$Mo values of NC meteorites are not only correlated with another, but also with $\varepsilon^{50}$Ti, $\varepsilon^{54}$Cr, and $\varepsilon^{62}$Ni (Fig. 3). These correlations involve lithophile (Ti, Cr) and siderophile (Ni, Mo) as well as refractory (Ti, Mo) and non-refractory (Cr, Ni) elements, indicating that the isotopic variations do not reflect the heterogeneous distribution of individual presolar carriers (e.g., SiC) or chemically fractionated components (e.g., refractory inclusions, silicates, metal). Instead, they are indicative of mixing between two isotopically distinct components with similar bulk chemical compositions. One of the mixing endmembers has the characteristic isotopic composition of the NC reservoir (low $\Delta^{95}$Mo, $\varepsilon^{50}$Ti, $\varepsilon^{54}$Cr, and $\varepsilon^{62}$Ni), while the other has high $\Delta^{95}$Mo, $\varepsilon^{50}$Ti, $\varepsilon^{54}$Cr, and $\varepsilon^{62}$Ni (Fig. 3a-c), which are the isotopic characteristics of bulk CC meteorites and Ca-Al-rich inclusions (CAIs).

However, unlike for $\Delta^{95}$Mo (Fig. 3a-c), NC meteorites, CC meteorites, and CAIs do not define a single mixing line in $\varepsilon^{94}$Mo versus $\varepsilon^{50}$Ti–$\varepsilon^{54}$Cr–$\varepsilon^{62}$Ni diagrams (Fig. 3d-f). Instead, NC meteorites plot along a trend towards more positive $\varepsilon^{50}$Ti, $\varepsilon^{54}$Cr, and $\varepsilon^{62}$Ni, but negative $\varepsilon^{94}$Mo (Fig. 3d-f). By contrast, bulk CC meteorites and typical CAIs are characterized by positive $\varepsilon^{94}$Mo and, therefore, plot off this trend (Fig. 3d-f). This also includes CI chondrites, which have been suggested to represent the material that was added to the inner disk and continuously changed its composition (Schiller et al. 2020). Thus, although one of the endmembers defining the NC mixing trend has some isotopic characteristics of CC meteorites and CAIs, compared to these samples this material is characterized by negative $\varepsilon^{94}$Mo, which is indicative of an excess in *s*-process Mo. The only known meteoritic materials with such a composition are the matrix of the CV3 chondrite Allende (Budde et al. 2016) and some fine-grained CAIs (Brennecka et al. 2017). We emphasize that this does not imply that these materials physically represent one of the endmembers defining the NC mixing trend, but it merely reveals that material with appropriate isotopic compositions existed in the disk at various times.

Like the NC mixing trend, the NC-CC dichotomy probably also results from mixing between two reservoirs with overall chondritic chemical but distinct isotopic compositions (Burkhardt et al. 2019; Nanne et al. 2019). In this model, the earliest disk, which formed by viscous spreading of early infalling material (Jacquet et al. 2019; Yang & Ciesla 2012), was characterized by a CAI-like isotopic composition (termed IC for Inclusion-like Chondritic reservoir; Burkhardt et al. 2019), while the later infall had a NC-like isotopic composition and provided most of the material in the inner disk. Mixing within the disk then gave rise to the CC reservoir, whose isotopic composition is intermediate between those of the IC and NC reservoirs (Burkhardt et al. 2019; Nanne et al. 2019). Thus, similar mixing processes that produced the NC-CC dichotomy also seem to be responsible for the isotopic variations within the NC reservoir, with the important difference that the material that produced the NC mixing trend is enriched in *s*-process Mo compared to the material that produced the NC-CC dichotomy (Fig. 3). Consequently, to account for both the isotopic variations in the NC reservoir and the NC-CC dichotomy requires at least three components: (1) the characteristic starting composition of the NC reservoir (e.g., as given by magmatic irons); (2) *s*-process-depleted IC material (as observed for most CAIs); and (3) *s*-process-enriched IC or CC material. Mixing between the first two of these components (i.e., between NC and IC) resulted in the characteristic composition of the CC reservoir, whereas mixing between the first and the third component (i.e., between NC and *s*-enriched IC or CC) produced the isotopic variations within the NC reservoir.

### 3.2. *Spatial and Temporal Variations in the NC Reservoir*

The addition of *s*-process-enriched IC or CC material to the inner disk may have occurred by different processes and at different times. For instance, isotopic heterogeneities in the inner disk may be inherited from the molecular cloud and would then reflect the changing isotopic composition of infalling matter from IC to NC at a very early stage when the NC reservoir was still forming (Burkhardt et al. 2019; Jacquet et al. 2019; Nanne et al. 2019). Alternatively, the outward transport of isotopically anomalous refractory material (e.g., CAIs) through the inner disk may have led to isotopic heterogeneities, because the fraction of CAIs remaining in the inner disk is expected to be higher at early times (Desch et al. 2018). Finally, the isotopic composition of the inner disk may have changed over time through the addition of CC-like dust from the outer Solar System after the NC-CC dichotomy had been established (Schiller et al. 2018). We note, however, that the NC mixing trend points towards *s*-enriched IC or CC material, rather than to the *s*-depleted IC or CC compositions as sampled by typical CAIs and bulk CC meteorites, respectively (Fig. 3). Thus, the NC mixing trend cannot result from the addition of these latter materials, but it may still reflect the addition of *s*-enriched IC or CC material. These additions would have likely resulted in distinct isotopic compositions for early- and late-accreted NC planetesimals, and so assessing whether the isotopic variations among NC meteorites are correlated with the accretion ages of their parent bodies may help identifying the underlying mechanisms that produced the NC mixing trend.

The accretion ages of NC iron meteorite and chondrite parent bodies are reasonably well established. For instance, $^{182}$Hf-$^{182}$W ages for most NC iron meteorites (i.e., group IC, IIAB, IIIAB, and IVA irons) indicate parent body accretion within <0.5 Ma after CAI formation (Kruijer et al. 2017, 2014). Only the IAB and IIE iron meteorite parent bodies may have accreted slightly later (Hunt et al. 2018; Kruijer & Kleine 2019), but their younger Hf-W ages may also reflect resetting during impact events, in which case the original accretion age is unknown (Kruijer & Kleine 2019). The parent bodies of NC chondrites accreted at ~2 Ma after CAI formation (Blackburn et al. 2017; Hellmann et al. 2019; Pape et al. 2019) and, therefore, later than those of the irons. Although accretion ages are only available for ordinary chondrites, it is reasonable to assume that the enstatite and Rumuruti chondrite parent bodies accreted at about the same time, given that all these bodies remained unmelted and, therefore, accreted later than ~1.5 Ma after CAI formation to avoid melting by $^{26}$Al decay (Hevey & Sanders 2006). Thus, when only iron meteorites and chondrites are considered, a temporal trend appears to exist in the isotopic composition of NC meteorites from lower to higher values of $\Delta^{95}$Mo, $\varepsilon^{50}$Ti, $\varepsilon^{54}$Cr, and $\varepsilon^{62}$Ni (Fig. 4). This trend is opposite to the expected isotopic variations for the incorporation of different amounts of refractory inclusions (e.g., CAIs) in NC planetesimals, which predicts more elevated $\Delta^{95}$Mo, $\varepsilon^{50}$Ti, $\varepsilon^{54}$Cr, and $\varepsilon^{62}$Ni in early-formed objects, reflecting the larger fraction of refractory inclusions in the inner disk at early times (Desch et al. 2018). Instead, the apparent temporal trend defined by iron meteorites and chondrites appears consistent with the expected effects of CC-dust addition to the inner disk, which should have produced more CC-like isotopic compositions in later-accreted NC bodies.

However, some other NC meteorites do not seem to fit the trend of isotope anomalies versus accretion ages very well. For instance, the ureilite parent body may have accreted as late as 1.5 Ma after CAIs (Budde et al. 2015), yet seems to have the lowest contribution of CC material among all NC meteorites. Moreover, acapulcoites-lodranites, whose parent body likely accreted at ~1.5 Ma after CAIs (Touboul et al. 2009), are also characterized by lower $\Delta^{95}$Mo, $\varepsilon^{50}$Ti, and $\varepsilon^{54}$Cr values than, for instance, the angrites, whose parent body likely accreted within the first ~0.5–1 Ma of the Solar System (Hans et al. 2013; Kleine et al. 2012). Finally, aubrites and enstatite chondrites have very similar isotopic compositions, but the aubrite parent body likely accreted earlier, well within ~1.5 Ma after CAI formation (Sugiura & Fujiya 2014). Differentiated meteorites, therefore, appear to cover most of the isotopic range observed among NC meteorites, yet these meteorites probably derive from bodies that accreted rather early. Similarly, later-accreted NC planetesimals (e.g., parent bodies of acapulcoites-lodranites and enstatite, ordinary, and Rumuruti chondrites) appear to cover a similar range of isotopic anomalies (Fig. 4). Together, these observations reveal that the NC isotopic mixing trend cannot solely reflect a temporal evolution of inner disk composition by addition of *s*-enriched CC dust from the outer Solar System.

The lack of a clear temporal trend in the inner disk's isotopic composition suggests that the NC mixing trend at least partially reflects spatial variations. These are unlikely to result from mixing between NC and CC materials, because this would, as noted above, lead to a temporal trend in the isotope anomalies. Instead, spatial variations within the inner disk more likely result from mixing between *s*-enriched IC and NC material, which occurred during infall from the Sun's parental molecular cloud and the associated early stages of disk building. It has been shown theoretically that infall from an isotopically zoned molecular cloud may result not only in an isotopically distinct outer disk (i.e., the CC reservoir), but also in spatial isotopic heterogeneities within the inner disk (Jacquet et al. 2019). The NC mixing trend may, therefore, at least in part reflect mixing of *s*-enriched IC and NC materials during infall and the early stages of disk building.

## 4. IMPLICATIONS FOR PLANETESIMAL FORMATION IN THE INNER DISK

As noted in prior studies, the clear compositional gap between the NC and CC reservoirs in multi-element isotope space (Fig. 3) requires the efficient separation of both reservoirs by a physical barrier, which may either be Jupiter itself (Kruijer et al. 2017) or, more generally, a pressure maximum in the disk (Brasser & Mojzsis 2020). This efficient separation implies that there has been only limited replenishment of dust in the inner disk through inward drifting CC dust. Thus, the inner disk is expected to become rapidly depleted in dust through rapid accretion into planetesimals (e.g., NC iron parent bodies) and loss to the Sun. This raises the question of how there was sufficient dust available in the inner disk for the ~2 Ma period of planetesimal formation inferred from the chronology of NC meteorites. Moreover, as noted above, if the NC isotopic mixing trend reflects spatial heterogeneities within the inner disk, then these isotopic variations must also be preserved for the ~2 Ma period of NC planetesimal formation. Together, these observations imply either that dust in the inner disk was somehow stored for at least ~2 Ma, or that later-formed NC planetesimals predominantly accreted from secondary dust produced by collisions among pre-existing planetesimals.

Pressure maxima in the inner disk are a potential way for storage of dust and would have also prevented mixing of dust across the resulting gap. However, a pressure bump would have also resulted in dust pile-up and, ultimately, its rapid accretion into planetesimals (e.g., Morbidelli et al. 2020). As such, it is unclear why some of these putative pressure maxima in the inner disk would have converted dust into planetesimals very rapidly (e.g., NC iron parent bodies), while others preserved dust for ~2 Ma until accretion into planetesimals (e.g., NC chondrite parent bodies). This would require different efficiencies with which pressure maxima resulted in the concentration of dust, but whether this is feasible is unknown. Thus, although we cannot exclude that pressure maxima in the inner disk resulted in a prolonged preservation of dust, the distinct accretion times of NC meteorite parent bodies make this scenario less likely.

By contrast, secondary dust would be produced naturally in the inner disk during the later stages of its evolution, when the damping effect of gas on the planetesimals' velocity dispersion becomes weaker and protoplanets become more massive so that they can scatter planetesimals more efficiently (Gerbig et al. 2019). Moreover, the lower amount of gas remaining at later stages favors planetesimal formation by the streaming instability, because the dust-to-gas ratio is high even for low amounts of dust (Carrera et al. 2017). Thus, from a dynamical standpoint the formation of planetesimals from secondary dust is expected, and so we consider it the more likely mechanism to account for the prolonged interval of planetesimal formation in the inner disk.

NC chondrites have broadly solar iron-to-metal ratios and overall chondritic relative abundances of non-volatile elements, indicating formation from chemically unfractionated dust (Palme et al. 2014). Thus, forming NC chondrites from collisionally-produced dust requires that this dust predominantly derives from small planetesimals that were unable to chemically differentiate, or from the primitive crust of larger, differentiated objects (Elkins-Tanton et al. 2011). Alternatively, highly energetic collisions may have resulted in vaporization of the colliding planetesimals, as has been suggested in some recent models for chondrule formation in the inner Solar System (Lock et al. 2019; Stewart et al. 2019). We note, however, that the formation of NC chondrites from collisionally-produced dust does not necessarily imply that the chondrules themselves formed as a result of these collisions. It is also possible that the chondrule-melting events occurred later by another process, and were unrelated to the collisions that produced their precursor dust. Distinguishing between these different models is not possible using the data of this study, but will require a better understanding of the underlying mechanisms that produced chondrules and whether or not this process was different in the inner and outer Solar System.

Finally, the formation of NC chondrites from secondary dust implies that their isotopic composition does not provide a snapshot of inner disk composition at the time of parent body accretion at ~2 Ma, but instead reflects those of pre-existing planetesimals and, therefore, records an earlier time of disk evolution. As such, there is no need to preserve spatial isotopic variations within the NC reservoir for a period of ~2 Ma. Instead, the isotopic variations among NC meteorites were likely generated over a much shorter time interval and, as such, may record a rapidly changing composition of the disk, where each planetesimal locks the instant composition of the disk when it forms.

## ACKNOWLEDGMENTS

Constructive reviews by A. Treiman and an anonymous reviewer greatly improved this paper and are gratefully acknowledged. We also thank E. G. Rivera-Valentín for efficient editorial handling, and the American Museum of Natural History (New York City), the Field Museum of Natural History (Chicago), the National History Museum (London), and the Smithsonian Institution (Washington, DC) for generously providing meteorite samples for this study. This project was funded by the Deutsche Forschungsgemeinschaft (DFG, German Research Foundation) – Project-ID 263649064 – TRR 170. This is TRR 170 pub. no. 102.

**Table 1**
Mo and Pt isotope data for IC, IIAB, IID, IIIAB, and IIIE iron meteorite groups.

| Sample | $N^a$ (Mo-IC) | $N^a$ (Pt-IC) | $\varepsilon^{92}Mo_{meas.}$ (± 95% CI) | $\varepsilon^{94}Mo_{meas.}$ (± 95% CI) | $\varepsilon^{95}Mo_{meas.}$ (± 95% CI) | $\varepsilon^{97}Mo_{meas.}$ (± 95% CI) | $\varepsilon^{100}Mo_{meas.}$ (± 95% CI) | $\Delta^{95}Mo$ (± 95% CI) | $\varepsilon^{192}Pt$ (± 95% CI) | $\varepsilon^{194}Pt$ (± 95% CI) | $\varepsilon^{196}Pt$ (± 95% CI) | References |
|---|---|---|---|---|---|---|---|---|---|---|---|---|
| **IC iron meteorites** | | | | | | | | | | | | |
| Chihuahua City | 6 | 1 | 0.92 ± 0.08 | 0.88 ± 0.06 | 0.41 ± 0.04 | 0.25 ± 0.06 | 0.21 ± 0.06 | | 0.06 ± 1.14 | 0.21 ± 0.15 | 0.09 ± 0.07 | this study |
| Mt. Dooling | 6 | 5 | 1.00 ± 0.13 | 0.91 ± 0.08 | 0.39 ± 0.04 | 0.25 ± 0.03 | 0.18 ± 0.05 | | 0.10 ± 0.47 | -0.01 ± 0.04 | -0.01 ± 0.04 | this study |
| Arispe | 5 | 3 | 0.77 ± 0.20 | 0.75 ± 0.14 | 0.21 ± 0.10 | 0.14 ± 0.07 | 0.27 ± 0.07 | | 13.69 ± 1.30 | 0.67 ± 0.11 | 0.42 ± 0.07 | 1,2 |
| Bendego | 5 | 7 | 0.83 ± 0.07 | 0.83 ± 0.13 | 0.26 ± 0.06 | 0.23 ± 0.11 | 0.31 ± 0.18 | | 0.79 ± 0.78 | 0.36 ± 0.05 | 0.48 ± 0.05 | 1,2 |
| **IC (int.-der.)[b]** | | | **0.96 ± 0.08** | **0.90 ± 0.06** | **0.40 ± 0.03** | **0.25 ± 0.03** | **0.19 ± 0.04** | **-14 ± 5** | | | | |
| **IC (indiv.-corr.)[c]** | | | 0.99 ± 0.07 | 0.92 ± 0.06 | 0.41 ± 0.06 | 0.24 ± 0.07 | 0.21 ± 0.05 | -14 ± 7 | | | | |
| **IIAB iron meteorites** | | | | | | | | | | | | |
| Ainsworth | 8 | 1 | 0.80 ± 0.07 | 0.82 ± 0.07 | 0.10 ± 0.05 | 0.23 ± 0.05 | 0.45 ± 0.07 | | 0.80 ± 1.14 | 0.57 ± 0.15 | 1.09 ± 0.06 | this study |
| Braunau | 7 | 5 | 1.40 ± 0.11 | 1.21 ± 0.08 | 0.58 ± 0.07 | 0.34 ± 0.08 | 0.29 ± 0.09 | | 0.28 ± 0.75 | -0.01 ± 0.05 | -0.03 ± 0.06 | this study |
| Guadalupe y Calvo | 5 | 5 | 1.27 ± 0.27 | 1.14 ± 0.12 | 0.51 ± 0.07 | 0.33 ± 0.08 | 0.32 ± 0.14 | | 0.36 ± 0.75 | -0.05 ± 0.05 | -0.05 ± 0.06 | this study |
| Mount Joy | 7 | 5 | 1.27 ± 0.06 | 1.08 ± 0.08 | 0.46 ± 0.06 | 0.28 ± 0.05 | 0.37 ± 0.05 | | 0.25 ± 0.57 | 0.18 ± 0.05 | 0.26 ± 0.04 | this study |
| North Chile | 5 | 5 | 1.29 ± 0.13 | 1.13 ± 0.08 | 0.50 ± 0.05 | 0.28 ± 0.06 | 0.34 ± 0.13 | | 0.03 ± 1.53 | 0.02 ± 0.04 | 0.02 ± 0.03 | this study |
| Sikhote Alin | 7 | 1 | 1.12 ± 0.16 | 1.03 ± 0.11 | 0.40 ± 0.08 | 0.24 ± 0.08 | 0.39 ± 0.09 | | 0.67 ± 1.14 | 0.32 ± 0.11 | 0.32 ± 0.09 | this study |
| **IIAB (int.-der.)[b]** | | | **1.37 ± 0.06** | **1.16 ± 0.04** | **0.53 ± 0.03** | **0.30 ± 0.03** | **0.33 ± 0.05** | **-16 ± 4** | | | | |
| **IIAB (indiv.-corr.)[c]** | | | 1.32 ± 0.06 | 1.15 ± 0.03 | 0.53 ± 0.03 | 0.31 ± 0.03 | 0.33 ± 0.02 | -15 ± 4 | | | | |
| **IID iron meteorites** | | | | | | | | | | | | |
| Carbo | 8 | 7 | 1.28 ± 0.12 | 0.94 ± 0.10 | 0.64 ± 0.09 | 0.42 ± 0.04 | 0.68 ± 0.08 | | 33.57 ± 0.40 | 1.27 ± 0.07 | 0.79 ± 0.04 | this study |
| Rodeo | 6 | 5 | 1.63 ± 0.16 | 1.19 ± 0.09 | 0.98 ± 0.09 | 0.51 ± 0.03 | 0.52 ± 0.12 | | -0.01 ± 0.65 | -0.01 ± 0.08 | -0.02 ± 0.03 | this study |
| Bridgewater | 7 | 5 | 1.63 ± 0.10 | 1.16 ± 0.16 | 0.96 ± 0.15 | 0.51 ± 0.12 | 0.67 ± 0.17 | | 0.80 ± 0.90 | 0.02 ± 0.08 | -0.01 ± 0.02 | 1,2 |
| N'kandhla | 5 | 5 | 1.71 ± 0.15 | 1.20 ± 0.14 | 1.02 ± 0.03 | 0.50 ± 0.03 | 0.59 ± 0.07 | | 0.64 ± 0.23 | 0.03 ± 0.05 | 0.01 ± 0.05 | 1,2 |
| **IID (int.-der.)[b]** | | | **1.65 ± 0.07** | **1.18 ± 0.07** | **1.01 ± 0.03** | **0.50 ± 0.02** | **0.58 ± 0.06** | **31 ± 5** | | | | |
| **IID (indiv.-corr.)[c]** | | | 1.66 ± 0.06 | 1.18 ± 0.03 | 0.98 ± 0.06 | 0.50 ± 0.02 | 0.59 ± 0.10 | 27 ± 6 | | | | |
| **IIIAB iron meteorites** | | | | | | | | | | | | |
| Boxhole | 9 | 9 | 0.98 ± 0.05 | 0.89 ± 0.06 | 0.35 ± 0.02 | 0.27 ± 0.04 | 0.30 ± 0.05 | | 23.16 ± 0.53 | 0.77 ± 0.04 | 0.41 ± 0.01 | this study |
| Cape York | 8 | 5 | 1.09 ± 0.12 | 1.01 ± 0.06 | 0.47 ± 0.07 | 0.23 ± 0.05 | 0.26 ± 0.09 | | -0.19 ± 0.57 | 0.05 ± 0.04 | 0.01 ± 0.04 | this study |
| Costilla Peak | 7 | 7 | 1.13 ± 0.14 | 1.03 ± 0.10 | 0.48 ± 0.09 | 0.27 ± 0.05 | 0.25 ± 0.10 | | -0.14 ± 0.73 | -0.02 ± 0.03 | -0.04 ± 0.03 | this study |
| Henbury | 7 | 8 | 1.05 ± 0.19 | 0.97 ± 0.10 | 0.42 ± 0.05 | 0.26 ± 0.02 | 0.17 ± 0.14 | | 15.58 ± 0.50 | 0.48 ± 0.04 | 0.26 ± 0.04 | this study |
| Willamette | 7 | 7 | 1.07 ± 0.16 | 0.92 ± 0.13 | 0.43 ± 0.07 | 0.25 ± 0.10 | 0.20 ± 0.08 | | -0.38 ± 0.66 | -0.02 ± 0.06 | -0.05 ± 0.04 | this study |
| Youanmi | 7 | 4 | 1.12 ± 0.05 | 1.02 ± 0.07 | 0.39 ± 0.06 | 0.25 ± 0.07 | 0.22 ± 0.06 | | 4.39 ± 0.48 | 0.45 ± 0.02 | 0.29 ± 0.09 | this study |
| **IIIAB (int.-der.)[b]** | | | **1.15 ± 0.07** | **1.01 ± 0.04** | **0.46 ± 0.04** | **0.26 ± 0.03** | **0.22 ± 0.05** | **-15 ± 5** | | | | |
| **IIIAB (indiv.-corr.)[c]** | | | 1.15 ± 0.07 | 1.02 ± 0.06 | 0.48 ± 0.03 | 0.27 ± 0.03 | 0.21 ± 0.05 | -13 ± 5 | | | | |
| **IIIE (indiv.-corr.)[c]** | | | 1.08 ± 0.06 | 0.96 ± 0.02 | 0.46 ± 0.06 | 0.30 ± 0.04 | 0.27 ± 0.09 | -11 ± 6 | | | | 1,2 |

**Notes.** The Mo and Pt isotope ratios were normalized to $^{98}Mo/^{96}Mo = 1.453173$ and $^{198}Pt/^{195}Pt = 0.2145$ using the exponential law, respectively. The $\varepsilon$-notation is the parts per $10^4$ deviation relative to the terrestrial bracketing Alfa Aesar solution standard. The uncertainties for $N \leq 3$ represent the 2 standard deviations (2 s.d.) of repeated analyses of the NIST 129c metal standard or the internal precision (2 standard errors [2 s.e.]), whichever is larger. The uncertainties for $N \geq 4$ represent the Student-t 95% confidence intervals, *i.e.*, $(t_{0.95}, N-1 \times s.d.)/\sqrt{N}$.
[a] Number of measurements.
[b] Intercept-derived values at $\varepsilon^{196}Pt = 0$ from $\varepsilon^iMo$–$\varepsilon^{196}Pt$ correlations for each group.
[c] Calculated using the weighted average $\varepsilon^iMo$–$\varepsilon^{196}Pt$ slopes determined for the IC, IIAB, IID, and IIIAB iron groups (-0.46±0.14 for $\varepsilon^{92}Mo$, -0.296±0.059 for $\varepsilon^{94}Mo$, -0.37±0.12 for $\varepsilon^{95}Mo$, -0.081±0.084 for $\varepsilon^{97}Mo$, and 0.130±0.058 for $\varepsilon^{100}Mo$) and the measured $\varepsilon^{196}Pt$.
**References.** (1) Worsham et al. (2019); (2) Kruijer et al. (2017)

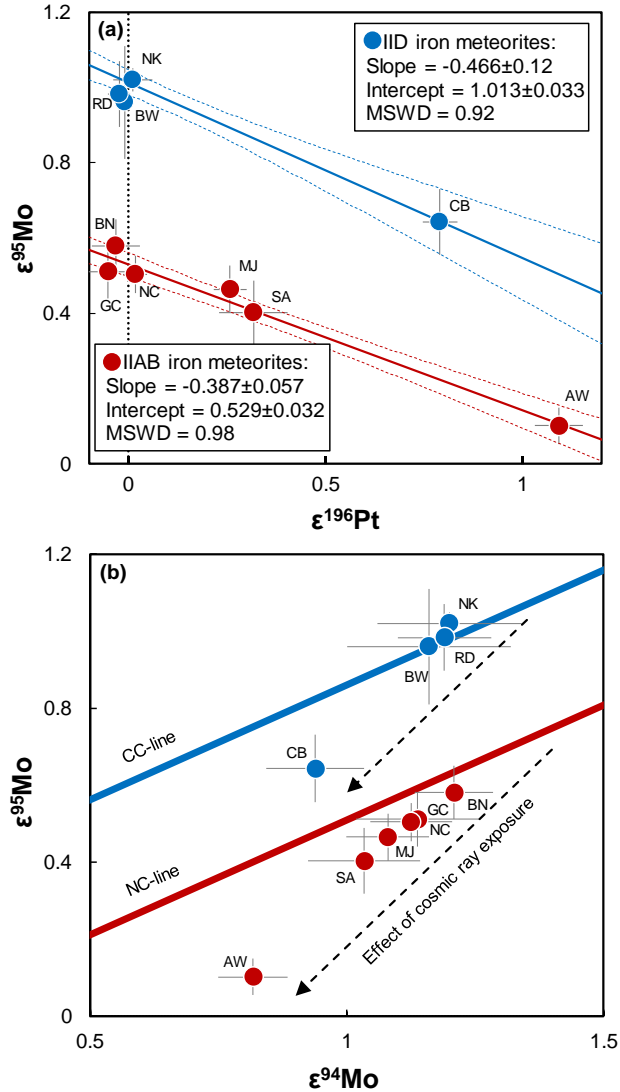

**Figure 1.** (a) $\varepsilon^{95}$Mo versus $\varepsilon^{196}$Pt for IIAB and IID iron meteorites. Both iron groups define precise and parallel correlation lines. Similar correlations are obtained for the other Mo isotopes as well as for the IC and IIIAB iron meteorites. The CRE-effects on $\varepsilon^{196}$Pt are predominantly governed by the reaction $^{195}$Pt(n,γ)$^{196}$Pt and the comparably large neutron capture cross section and resonance integral for $^{195}$Pt (Mughabghab 2003). For Mo isotopes the most important neutron capture reaction is $^{95}$Mo(n,γ)$^{96}$Mo, and since $^{96}$Mo is used as normalizing isotope, any CRE-effect on $^{96}$Mo is transposed into all $\varepsilon^{i}$Mo values. (b) $\varepsilon^{95}$Mo versus $\varepsilon^{94}$Mo for IIAB and IID irons, showing that unaccounted CRE-effects can result in significant departure from the NC- and CC-lines as defined in Budde et al. (2019). AW: Ainsworth, BN: Braunau, BW: Bridgewater, CB: Carbo, GC: Guadalupe y Calvo, MJ: Mount Joy, NC: North Chile, NK: N'kandhla, RD: Rodeo, SA: Sikhote Alin.

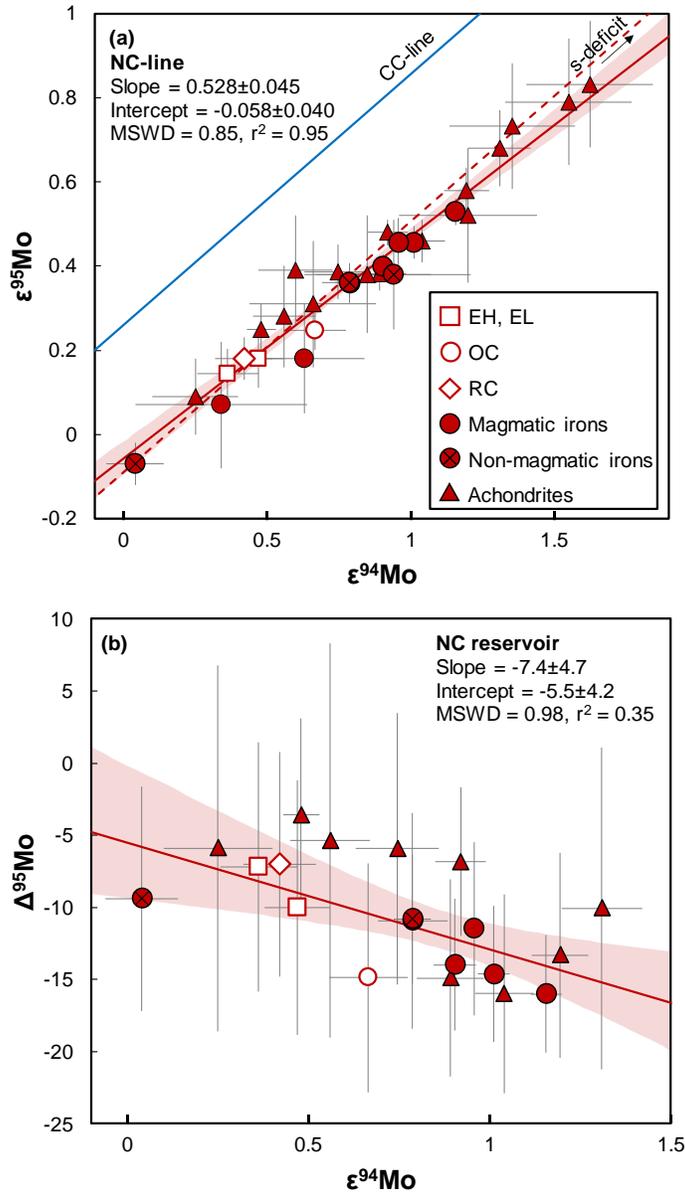

**Figure 2.** $\varepsilon^{95}$Mo versus $\varepsilon^{94}$Mo (a) and $\Delta^{95}$Mo versus $\varepsilon^{94}$Mo (b) for NC meteorites. CC-line (blue) and NC-line (dashed red line) as defined in Budde et al. (2019) shown for reference. All regressions were calculated using the model 1 fit of Isoplot (Ludwig 2003). (a) Note that some iron meteorites plot below the previously defined NC-line and that NC meteorites plot along a line with a slightly shallower slope compared to the CC-line. (b) NC meteorites define a weak inverse correlation of $\Delta^{95}$Mo versus $\varepsilon^{94}$Mo (only includes samples having $\sigma_{\Delta 95Mo}$ < 15 ppm). Correlated uncertainties for $\Delta^{95}$Mo and $\varepsilon^{94}$Mo were taken into account in the regression, but error ellipses are omitted for clarity.

Isotopic evolution of the inner Solar System

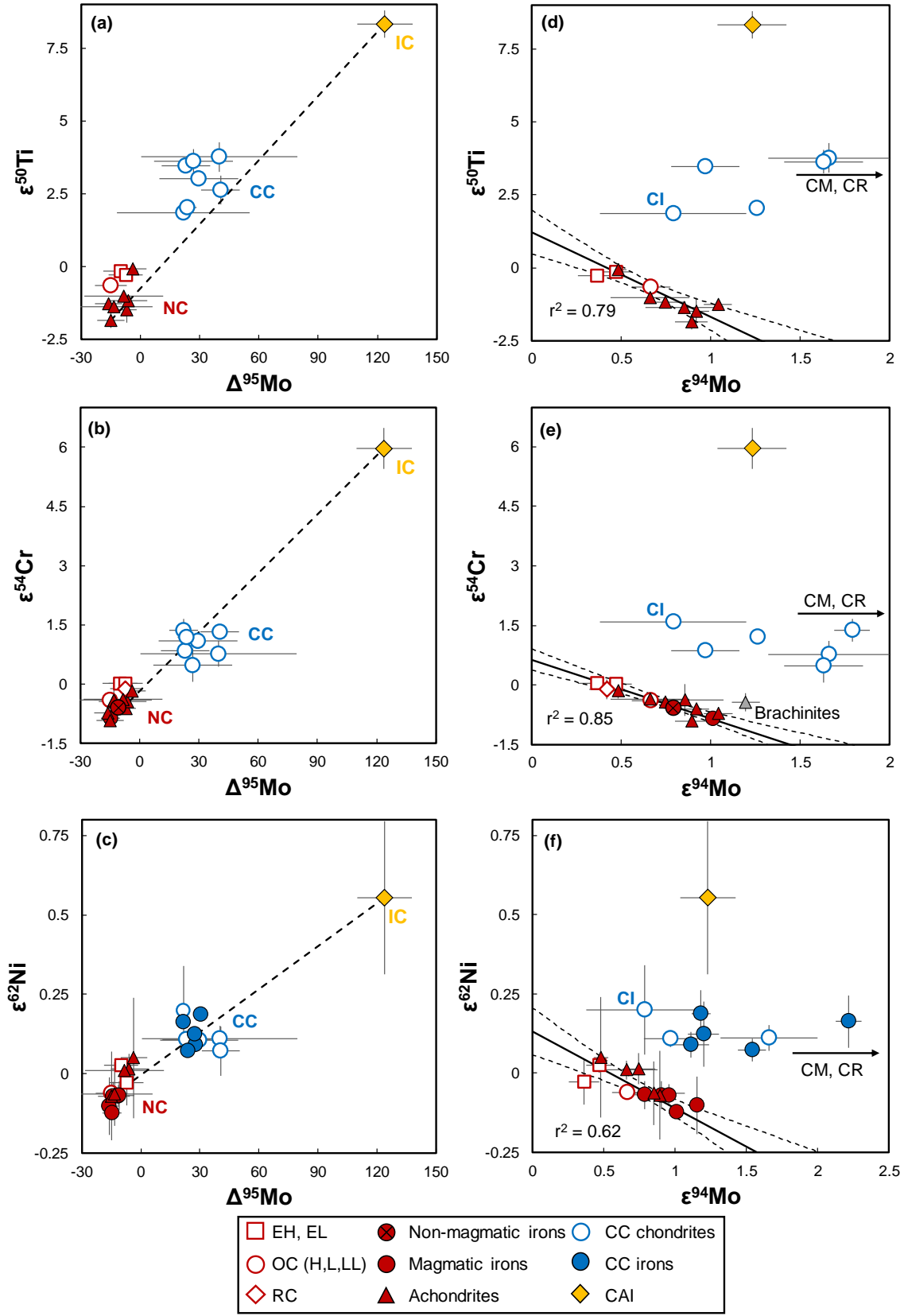

Isotopic evolution of the inner Solar System**Figure 3.** $\Delta^{95}$Mo and $\varepsilon^{94}$Mo versus $\varepsilon^{50}$Ti, $\varepsilon^{54}$Cr, and $\varepsilon^{62}$Ni for NC meteorites, CC meteorites, and CAI (labelled 'IC' for Inclusion-like Chondritic reservoir). For data sources see Table S1. (a-c) For $\Delta^{95}$Mo versus $\varepsilon^{50}$Ti–$\varepsilon^{54}$Cr–$\varepsilon^{62}$Ni, the composition of all meteorites can be accounted for by mixing between an initial NC reservoir characterized by the lowest $\Delta^{95}$Mo–$\varepsilon^{50}$Ti–$\varepsilon^{54}$Cr–$\varepsilon^{62}$Ni values and the IC reservoir, as indicated by the dashed black line. (d-f) For $\varepsilon^{94}$Mo versus $\varepsilon^{50}$Ti–$\varepsilon^{54}$Cr–$\varepsilon^{62}$Ni, only NC meteorites display correlated variations, but bulk CC meteorites and CAI plot off these trends towards more positive $\varepsilon^{94}$Mo (i.e., *s*-process depleted compositions). Note that CM and CR chondrites plot off scale towards larger $\varepsilon^{94}$Mo. Solid black line is a linear regression through data for NC meteorites, calculated using Isoplot (Ludwig 2003). Dashed lines show the error envelope of the regression. Note that brachinites are excluded from the regression, because their isotopic composition might have been modified during partial differentiation (Hopp et al. 2020).

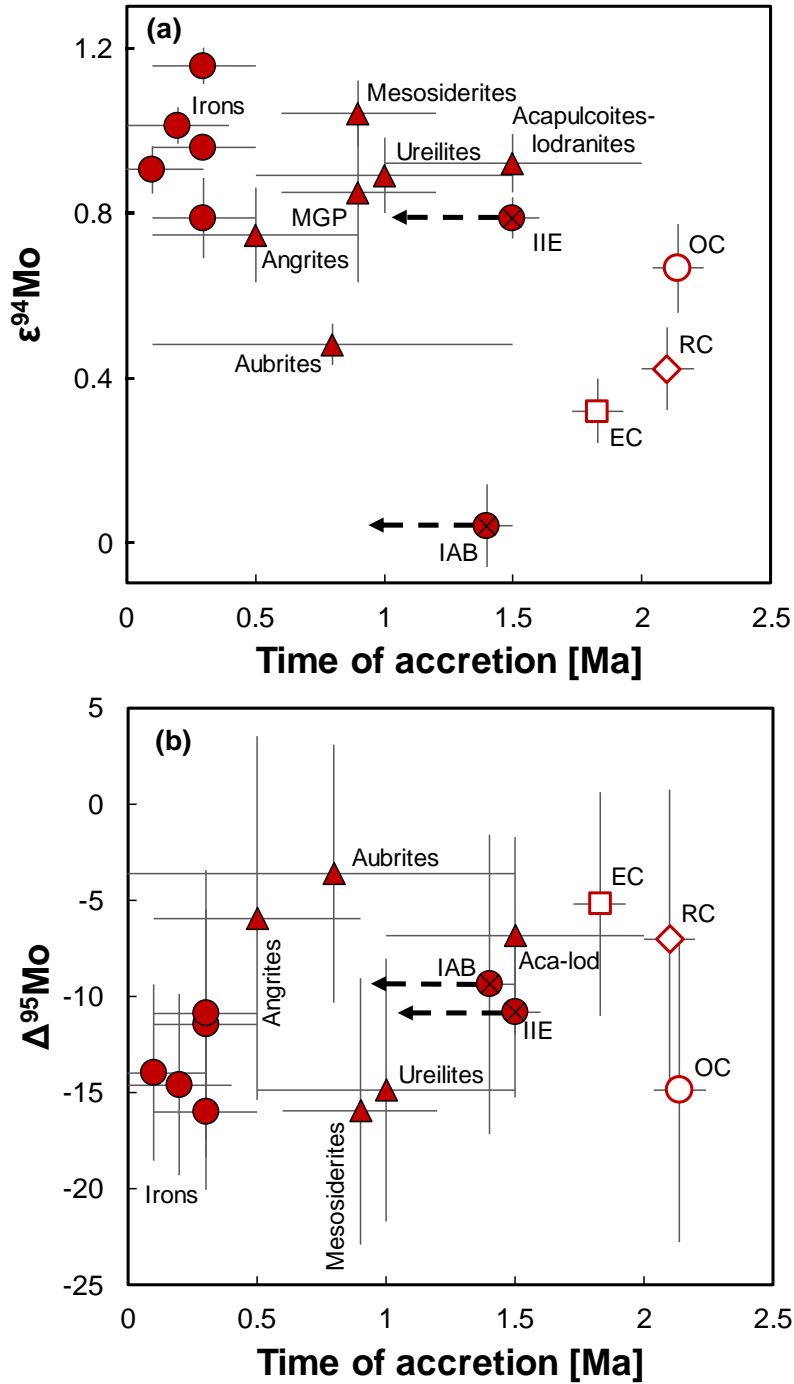

**Figure 4.** $\varepsilon^{94}$Mo (a) and $\Delta^{95}$Mo (b) versus accretion ages for NC meteorites. Accretion ages are summarized in table S1. Note that some anomalous ureilites display larger $\varepsilon^{94}$Mo anomalies but these samples are not shown here because their accretion ages are unknown (see table S1). Arrows indicate that accretion ages may be older than shown. Legend as in Fig. 3. MGP: Main Group Pallasites.

# APPENDIX

**Table S1**
Summary of Mo, Ti, Cr, Ni, and accretion age literature data for selected meteorites

| Sample | $\varepsilon^{94}$Mo | 95% CI | $\varepsilon^{95}$Mo | 95% CI | $\Delta^{95}$Mo | 95% CI | rho | references | ← comment | $\varepsilon^{50}$Ti | 95% CI | references | $\varepsilon^{54}$Cr | 95% CI | references | $\varepsilon^{62}$Ni | 95% CI | references | accretion age [Ma][a] | 2σ | references |
|---|---|---|---|---|---|---|---|---|---|---|---|---|---|---|---|---|---|---|---|---|---|
| *Non-carbonaceous (NC) meteorites* | | | | | | | | | | | | | | | | | | | | | |
| Chondrites | | | | | | | | | | | | | | | | | | | | | |
| EH | 0.47 | 0.09 | 0.18 | 0.07 | -10 | 9 | -0.61 | 1,2 | weighted mean | -0.14 | 0.07 | 18,19 | 0.02 | 0.05 | 28,29,30 | 0.03 | 0.03 | 51,52,53 | 1.83 | 0.10 | 58 |
| EL | 0.36 | 0.11 | 0.14 | 0.06 | -7 | 9 | -0.51 | 1,3 | simple mean + 95% CI | -0.28 | 0.17 | 3,19,20 | 0.03 | 0.06 | 28,29,30 | -0.03 | 0.07 | 3,51,52,53 | 1.83 | 0.10 | 58 |
| OC (H, L, LL) | 0.67 | 0.11 | 0.25 | 0.05 | -15 | 8 | -0.81 | 1,2,3,4 | simple mean + 95% CI[b] | -0.66 | 0.06 | 3,18,19,20,21,22 | -0.40 | 0.04 | 28,29,31 | -0.06 | 0.02 | 3,51,52,53,54 | 2.14 | 0.10 | 58 |
| RC | 0.42 | 0.10 | 0.18 | 0.05 | -7 | 8 | -0.77 | 5 | | | | | -0.11 | 0.25 | 32 | | | | 2.10 | 0.10 | 58 |
| Achondrites | | | | | | | | | | | | | | | | | | | | | |
| Mesosiderites | 1.04 | 0.08 | 0.46 | 0.05 | -16 | 7 | -0.69 | 5 | | -1.27 | 0.16 | 18 | -0.72 | 0.07 | 28 | | | | 0.90 | 0.30 | 58 |
| Acapulcoites-Lodranites | 0.92 | 0.07 | 0.48 | 0.03 | -7 | 5 | -0.84 | 10 | | -1.48 | 0.45 | 23 | -0.62 | 0.15 | 23,32,33,34 | | | | 1.50 | 0.50 | 63 |
| Winonaites | 0.25 | 0.15 | 0.09 | 0.09 | -6 | 13 | -0.70 | 10 | | | | | | | | | | | | | |
| Brachinites | 1.20 | 0.08 | 0.58 | 0.05 | -13 | 7 | -0.65 | 5,10 | weighted mean | | | | -0.44 | 0.23 | 36 | | | | | | |
| Ureilites | 0.89 | 0.09 | 0.38 | 0.04 | -15 | 7 | -0.80 | 5 | | -1.85 | 0.26 | 18 | -0.92 | 0.04 | 29,32,37,38 | -0.07 | 0.14 | 52 | 1.00 | 0.50 | 59,62 |
| Angrites | 0.75 | 0.11 | 0.39 | 0.06 | -6 | 9 | -0.72 | 5 | | -1.18 | 0.08 | 18,19 | -0.43 | 0.06 | 28,31,32,39 | 0.01 | 0.05 | 52 | 0.50 | 0.40 | 58,64,65 |
| Aubrites | 0.48 | 0.05 | 0.25 | 0.06 | -4 | 7 | -0.44 | 5 | | -0.06 | 0.11 | 19 | -0.16 | 0.19 | 28 | 0.05 | 0.19 | 52 | 0.80 | 0.70 | 58 |
| Main group pallasites | 0.85 | 0.22 | 0.38 | 0.14 | -13 | 19 | -0.68 | 7 | | -1.37 | 0.08 | 18 | -0.39 | 0.40 | 28,29 | -0.06 | 0.10 | 35 | 0.90 | 0.30 | 58 |
| EET 87517 (anom. Ureilite) | 1.62 | 0.22 | 0.83 | 0.15 | -14 | 20 | -0.66 | 5 | | | | | | | | | | | | | |
| PCA 82506 (anom. Ureilite) | 1.35 | 0.22 | 0.73 | 0.15 | -7 | 20 | -0.65 | 5 | | | | | | | | | | | | | |
| NWA 1058 (ungr.) | 1.31 | 0.11 | 0.68 | 0.09 | -10 | 11 | -0.59 | 5 | | | | | | | | | | | | | |
| NWA 6112 (ungr.) | 1.55 | 0.22 | 0.79 | 0.15 | -13 | 20 | -0.66 | 10 | | | | | | | | | | | | | |
| NWA 5363/5400 (ungr.) | 0.66 | 0.22 | 0.31 | 0.15 | -8 | 20 | -0.60 | 10 | | -1.02 | 0.10 | 3 | -0.37 | 0.13 | 3 | 0.01 | 0.03 | 3 | | | |
| NWA 2526 (ungr.) | 0.60 | 0.13 | 0.39 | 0.13 | 3 | 15 | -0.52 | 10 | | | | | | | | | | | | | |
| NWA 725 (anom. acap.) | 1.20 | 0.24 | 0.52 | 0.16 | -20 | 21 | -0.67 | 2 | | | | | | | | | | | | | |
| NWA 11048 (anom. acap.) | 0.56 | 0.11 | 0.28 | 0.12 | -5 | 14 | -0.48 | 10 | | | | | | | | | | | | | |
| Iron meteorites | | | | | | | | | | | | | | | | | | | | | |
| IC | 0.90 | 0.06 | 0.40 | 0.03 | -14 | 5 | -0.75 | this study, 6 | int.-der. | | | | | | | -0.07 | 0.04 | 51,55,56 | 0.10 | 0.20 | 14 |
| IIAB | 1.16 | 0.04 | 0.53 | 0.03 | -16 | 4 | -0.62 | this study | int.-der. | | | | | | | -0.10 | 0.09 | 51,52,55 | 0.30 | 0.20 | 14 |
| IIE | 0.79 | 0.05 | 0.36 | 0.03 | -11 | 4 | -0.68 | 8 | weighted mean | | | | -0.59 | 0.13 | 28 | | | | 1.50 | +0.1/-X | 60 |
| IIIAB | 1.01 | 0.04 | 0.46 | 0.04 | -15 | 5 | -0.56 | this study | int.-der. | | | | -0.85 | 0.06 | 28 | -0.12 | 0.02 | 51,52,55 | 0.20 | 0.20 | 14 |
| IIIE | 0.96 | 0.02 | 0.46 | 0.06 | -11 | 6 | -0.20 | 6,14 | indiv.-corr. | | | | | | | -0.07 | 0.04 | 56 | 0.30 | 0.20 | 14 |
| IVA | 0.79 | 0.10 | 0.36 | 0.05 | -11 | 7 | -0.79 | 8,9 | simple mean + 95% CI | | | | | | | -0.07 | 0.04 | 51,52,55,56 | 0.30 | 0.20 | 14 |
| IAB sH | 0.94 | 0.27 | 0.38 | 0.13 | -18 | 21 | -0.78 | 2 | | | | | | | | | | | | | |
| IAB MG-sLL | 0.04 | 0.10 | -0.07 | 0.05 | -9 | 8 | -0.77 | 9 | | | | | | | | | | | 1.40 | +0.1/-X | 61 |
| Gebel Kamil (ungr.) | 0.34 | 0.30 | 0.07 | 0.15 | -13 | 23 | -0.76 | 9 | | | | | | | | | | | | | |
| Mont Dieu (ungr.) | 0.63 | 0.21 | 0.18 | 0.13 | -20 | 18 | -0.69 | 4 | | | | | | | | | | | | | |
| *Carbonaceous (CC) meteorites* | | | | | | | | | | | | | | | | | | | | | |
| Chondrites | | | | | | | | | | | | | | | | | | | | | |
| CI | 0.79 | 0.41 | 0.69 | 0.23 | 22 | 34 | | 7 | | 1.85 | 0.12 | 18,19 | 1.59 | 0.06 | 28,29,31,32,40,41 | 0.20 | 0.14 | 51,53,54 | 3.60 | 0.50 | 59 |
| CM | 4.82 | 0.20 | 3.17 | 0.16 | 30 | 20 | | 7 | | 3.02 | 0.09 | 18,19,21 | 1.10 | 0.08 | 28,29,40,42 | 0.10 | 0.03 | 51,52,53,54 | 3.50 | +0.7/-0.5 | 58 |
| CO | 1.66 | 0.34 | 1.39 | 0.34 | 40 | 40 | | 11 | | 3.77 | 0.50 | 18,19 | 0.77 | 0.33 | 28,29,40 | 0.11 | 0.04 | 51,53 | 2.70 | 0.20 | 58 |
| CV | 0.97 | 0.19 | 0.81 | 0.05 | 23 | 12 | | 12 | | 3.47 | 0.19 | 3,18,19,20,21,24 | 0.86 | 0.08 | 3,28,29,40 | 0.11 | 0.03 | 3,51,52,53,54,56 | 2.70 | 0.30 | 59 |
| CK | 1.63 | 0.22 | 1.24 | 0.15 | 27 | 20 | | 5 | | 3.63 | 0.40 | 18,19 | 0.48 | 0.42 | 28,29 | | | | 2.60 | 0.20 | 58 |
| CR | 3.11 | 0.15 | 2.26 | 0.04 | 41 | 10 | | 13 | | 2.63 | 0.49 | 18,19,24 | 1.34 | 0.03 | 24,28,29 | 0.07 | 0.08 | 51 | 3.85 | 0.15 | 59 |
| CH | 1.79 | 0.10 | 1.29 | 0.04 | 22 | 7 | | 5 | | | | | 1.37 | 0.29 | 28 | | | | | | |
| CB | 1.26 | 0.04 | 0.99 | 0.04 | 24 | 5 | | 5 | | 2.04 | 0.07 | 18 | 1.20 | 0.09 | 28,40,43,44 | | | | | | |
| Iron meteorites | | | | | | | | | | | | | | | | | | | | | |
| IIC | 2.22 | 0.09 | 1.54 | 0.06 | 22 | 8 | | 14 | | | | | | | | 0.16 | 0.08 | 56 | 0.90 | +0.4/-0.2 | 14 |
| IID | 1.18 | 0.07 | 1.01 | 0.03 | 31 | 5 | | this study,6,14 | int.-der. | | | | | | | 0.19 | 0.08 | 56 | 0.90 | +0.4/-0.2 | 14 |
| IIF | 1.11 | 0.13 | 0.94 | 0.08 | 28 | 11 | | 14 | | | | | | | | 0.09 | 0.04 | 56 | 0.90 | +0.4/-0.2 | 14 |
| IIIF | 1.20 | 0.11 | 0.99 | 0.04 | 27 | 8 | | 14 | | | | | | | | 0.12 | 0.10 | 56 | 1.00 | 0.20 | 14 |
| IVB | 1.54 | 0.10 | 1.16 | 0.05 | 24 | 8 | | 4 | | | | | | | | 0.07 | 0.04 | 51,52,55 | 1.00 | 0.20 | 14 |
| *Inclusion-like (IC) reservoir* | | | | | | | | | | | | | | | | | | | | | |
| CAI | 1.23 | 0.19 | 1.97 | 0.08 | 124 | 14 | | 7,15,16 | simple mean + 95% CI | 8.33 | 0.47 | 25,26,27 | 5.97 | 0.52 | 28,45,46,47,48,49,50 | 0.55 | 0.24 | 57 | | | |

**Notes.** The ε-notation is the parts per $10^4$ deviation relative to the terrestrial bracketing solution standard. The uncertainties represent the 2 standard deviations (2 s.d.) for samples with N ≤ 3 and Student-t 95% confidence intervals, i.e., $(t_{0.95},N-1 \times s.d.)/\sqrt{N}$, for N ≥ 4.

[a] after the formation of Ca-Al-rich inclusions (CAI)

[b] excluding Richardton metal and Saint-Séverin

**References.** (1) Render et al. (2017); (2) Worsham et al. (2017); (3) Burkhardt et al. (2017); (4) Yokoyama et al. (2019); (5) Budde et al. (2019); (6) Worsham et al. (2019); (7) Burkhardt et al. (2011); (8) Poole et al. (2017); (9) Bermingham al. (2018); (10) Hopp et al. (2020); (11) Burkhardt et al. (2014); (12) Budde et al. (2016); (13) Budde et al. (2018); (14) Kruijer et al. (2017); (15) Brennecka et al. (2013); (16) Shollenberger et al. (2018); (17) Williams et al. (2016); (18) Trinquier et al. (2009); (19) Zhang et al. (2012); (20) Gerber et al. (2017); (21) Zhang et al. (2011); (22) Bischoff et al. (2019); (23) Goodrich et al. (2017); (24) Sanborn et al. (2019); (25) Torrano et al. (2019); (26) Davis et al. (2018); (27) Render et al. (2019); (28) Trinquier et al. (2007); (29) Qin et al. (2010); (30) Mougel et al. (2018); (31) Schiller et al. (2014); (32) Larsen et al. (2011); (33) Göpel and Birck (2010); (34) Li et al. (2018); (35) Dauphas et al. (2008); (36) Sanborn and Yin (2015); (37) Yamakawa et al. (2010); (38) Zhu et al. (2020); (39) Zhu et al. (2019); (40) Shukolyukov and Lugmair (2006); (41) Petitat et al. (2011); (42) Göpel et al. (2015); (43) Yamashita et al. (2010); (44) Yamashita et al. (2005); (45) Birck and Lugmair (1988); (46) Birck and Allègre (1988); (47) Papanastassiou (1986); (48) Bogdanovski et al. (2002); (49) Mercer et al. (2015); (50) Torrano et al. (2018); (51) Regelous et al. (2008); (52) Tang and Dauphas (2012); (53) Steele et al. (2012); (54) Tang and Dauphas (2014); (55) Steele et al. (2011); (56) Nanne et al. (2019); (57) Render et al. (2018); (58) Sugiura et al. (2014); (59) Desch et al. (2018); (60) Kruijer et al. (2019); (61) Hunt et al. (2018); (62) Budde et al. (2015); (63) Touboul et al. (2009); (64) Hans et al. (2013); (65) Kleine et al. (2012)

**APPENDIX REFERENCES**